\documentclass[aps,prd,twocolumn,superscriptaddress,nofootinbib,showpacs,10pt]{revtex4-1}
\usepackage[utf8]{inputenc}
\usepackage[english]{babel}
\usepackage{amsmath}
\usepackage{amsfonts}
\usepackage{amssymb}
\usepackage{graphics}
\usepackage{graphicx}
\usepackage[left=2cm,right=2cm,top=2cm,bottom=2cm]{geometry}
\usepackage[usenames,dvipsnames,svgnames]{xcolor}  

\usepackage{hyperref}   
\hypersetup{colorlinks=true,linkcolor=beamer@PRD, citecolor=beamer@PRD}
\definecolor{beamer@PRD}{RGB}{46,48,146}

\newcommand\myref[1]{\textcolor{beamer@PRD}{(}\ref{#1}\textcolor{beamer@PRD}{)}}

\begin{document}
\title{Probing noncommutative theories with quantum optical experiments}
\author{Sanjib Dey}\email{dey@iisermohali.ac.in}\email{sanjibdey4@gmail.com}
\affiliation{Department of Physics, Indian Institute of Science Education and  Research Mohali, \protect\\ Sector 81, SAS Nagar, Manauli 140306, India}
\author{Anha Bhat}\email{anhajan1@gmail.com}
\affiliation{Department of Metallurgical and Materials Engineering, National Institute \protect\\ of Technology, Srinagar 190006, India}
\author{Davood Momeni}\email{davood@squ.edu.om}
\affiliation{Department of Physics, College of Science, Sultan Qaboos University, P.O. Box 36, Alkhod 123, Oman}
\author{Mir Faizal}\email{mirfaizalmir@googlemail.com}
\affiliation{Irving K. Barber School of Arts and Sciences, University of British Columbia-Okanagan,\protect\\ 3333 University Way, Kelowna, British Columbia V1V 1V7, Canada}
\affiliation{Department of Physics and Astronomy, University of Lethbridge, Lethbridge, Alberta T1K 3M4, Canada}
\author{Ahmed Farag Ali}\email{ahmed.ali@fsc.bu.edu.eg}
\affiliation{Department of Physics, Faculty of Science, Benha University, Benha, 13518, Egypt}
\affiliation{Netherlands Institute for Advanced Study, Korte Spinhuissteeg 3, 1012 CG Amsterdam, Netherlands}
\author{Tarun Kumar Dey}\email{deytarunkumar2020@gmail.com}
\affiliation{Department of Mathematics, Sambhunath College, Labpur 731303, India}
\author{Atikur Rehman}\email{atikurrhmn@gmail.com}
\affiliation{Department of Metallurgical and Materials Engineering, National Institute \protect\\ of Technology, Srinagar 190006, India}
\begin{abstract}
One of the major difficulties of modern science underlies at the unification of general relativity and quantum mechanics. Different approaches towards such theory have been proposed. Noncommutative theories serve as the root of almost all such approaches. However, the identification of the appropriate passage to quantum gravity is suffering from the inadequacy of experimental techniques. It is beyond our ability to test the effects of quantum gravity thorough the available scattering experiments, as it is unattainable to probe such high energy scale at which the effects of quantum gravity appear. Here we propose an elegant alternative scheme to test such theories by detecting the deformations emerging from the noncommutative structures. Our protocol relies on the novelty of an opto-mechanical experimental setup where the information of the noncommutative oscillator is exchanged via the interaction with an optical pulse inside an optical cavity. We also demonstrate that our proposal is within the reach of current technology and, thus, it could uncover a feasible route towards the realization of quantum gravitational phenomena thorough a simple table-top experiment.
\end{abstract}


\maketitle
\section{Introduction} \label{sec1}
In spite of having several serious proposals for the quantization of general relativity, a fully consistent quantum theory of gravity is yet unexplored and remains the most important open challenge in modern science. The understanding of a possible route to such theory is extremely barred by the scarcity of experimentally accessible phenomena. It is argued that the fundamental concept of space-time is not wholly compatible with that of the standard version of quantum mechanics, but fits mostly with its noncommutative version \cite{Connes,Garay,Seiberg_Witten,Madore,Douglas, Szabo}. By now, many approaches towards quantum gravity support the idea of noncommutativity \cite{Douglas_Hull,Amelino,Gomis,Kowalski, Chaichian,Nicolini}. It also plays an important role in theories like causal set approach \cite{Alvarez-Gaume,Akofor}. Noncommutative theories have been used to explore radically new ideas in various other branches of modern physics too \cite{Bellissard,Gopakumar,Gamboa,Calmet}. In string theory the smallest available length scale is the string length beyond which it is not possible to probe the space-time, and such minimal observable length is a natural consequence of noncommutative theories. Numerous different versions of noncommutative structure have been proposed in the literature \cite{Kempf,Seiberg_Witten,Szabo,Das_Vagenas_PRL} starting from the Lorentz covariant version of Snyder \cite{Snyder}, whose aim was to resolve the problem of ultraviolet divergence. Further possibilities arise due to the breaking of the Lorentz covariance of the Snyder's algebra \cite{Gomis,Seiberg_Susskind,Ardalan,Banerjee}. Perhaps the simplest and most commonly studied version follows from Seiberg and Witten \cite{Seiberg_Witten}, who explored that the string theory can be realized as an effective field theory in noncommutative space with the simple commutation relation $[x^\mu,x^\nu]=i\theta^{\mu\nu}$, where $\theta^{\mu\nu}$ is a constant antisymmetric tensor. A more general form, which is familiar as the noncommutative phase space, has been studied extensively in the literature \cite{Langmann,Li,Geloun}, and is given by
\begin{alignat}{1}\label{NCOM}
[x_i,x_j]&=i\tilde{\theta}_{ij}=i\frac{c_{ij}}{\Lambda_1}, \quad [p_i,p_j]=i\tilde{\Omega}_{ij}=i\frac{d_{ij}}{\Lambda_2}, \\
[x_i,p_j]&=i\hbar\delta_{ij}\left(1+\frac{\tilde{\theta}_{ik}\tilde{\Omega}_{kj}}{4\hbar^2}\right)=i\hbar\delta_{ij}\left(1+\frac{c_{ik}d_{kj}}{4\hbar^2\Lambda_1\Lambda_2}\right), \notag
\end{alignat}
where $\tilde{\theta}_{ij}=\theta_{ij}\hbar/(m\omega),~\tilde{\Omega}_{ij}=\hbar m \omega\Omega_{ij}$ and $\theta,\Omega$ are $3\times 3$ constant antisymmetric matrices with $\theta_{ij}, \Omega_{ij}$ being dimensionless. Here, $c_{ij}$ and $d_{ij}$ have the same properties as $\theta_{ij}$ and $\Omega_{ij}$, respectively and, thus, $\Lambda_1$ and $\Lambda_2$ have the dimension of the inverse squared length and inverse squared momentum, respectively. Thus, $\Lambda_1$ is the length scale or energy scale (in natural units) at which the effects of noncommutative space time becomes relevant. There are plenty of articles dealing with the bounds of such energy scale $\Lambda_1$ or $\tilde{\theta}_{ij}$; see, for instance \cite{Carroll,Carlson,Anisimov,Calmet,Das_Vagenas_PRL,Szabo_Review,Prakash_Mitra,Moumni, Ghegal,Joby}, among which the best currently accepted bounds on $\tilde{\theta}_{ij}$ are approximately $10^{-8}~\text{GeV}^{-2}$. It is straight forward to compute the bounds on the conjugate parameter $\tilde{\Omega}_{ij}\sim \text{MeV}$ by using the generalized uncertainty relations associated with the commutation relations \myref{NCOM}. Nevertheless, by following the standard minimization technique \cite{Kempf,Dey_Fring_Gouba}, the minimal measurable length for the noncommutative algebra \myref{NCOM} consistent with the generalized uncertainty principle results to $(\Delta x_i)_{\text{min}}=l_P\sqrt{\theta_{ik}\Omega_{kj}/4}$, with $l_P$ being the Planck length. Therefore, the minimal length resulting from the current lowest bound on the parameter $\tilde{\theta}_{ij}$ is of the order of $l_P\sqrt{\theta_{ik}\Omega_{kj}}\sim l_P\times 10^{6}$, which is yet far from the Planck length $l_P$. Obviously, the Planck scale accuracy can be achieved while the parameters $\theta$ and $\Omega$ are constrained by the relation $\theta_{ik}\Omega_{kj}\sim 1$. Our proposal indicates that it is not only possible to probe the current lower bound, but also the desired Planck length ($l_P\approx 10^{-35}m$) or the Planck energy ($E_P\approx 10^{19} GeV$) at which the effects of quantum gravity are expected to become relevant.

It should be mentioned that in near future the high energy scattering experiment may lead us to the feasibility of probing such high energy scales. However, current experimental range is about $15$ orders of magnitude away from the Planck energy and, thus, at this moment it is not clear whether it will be possible to reach the desired accuracy at all. Astronomical observations have also failed to provide any promising evidence to study the quantum gravitational effects. Thus, it is worth exploring some alternative ways and, noncommutative theories have been very successful so far to circumvent the quantum gravity problem at least theoretically. This is the reason why it would be very exciting if we could prove the existence of noncommutativity in nature.  

The purpose of the present manuscript is to propose a fascinating experimental scheme that may provide an alternative way to reach the energy scale near to the Planck length and, thus, probe the noncommutative theories.  Our system allows to measure the deformation of the canonical commutation relations in a novel parameter regime, thereby reaching remarkable sensitivity in measuring the effects at Planck scale. However, in our approach it is not required to probe the Planck scale accuracy in position measurement. Instead, what we use is a quantum mechanical ancillary system that can provide a way to measure any deformation of the canonical commutator directly. More precisely, we allow a massive noncommutative oscillator to interact with a strong optical field in an opto-mechanical cavity by utilizing the principle of radiation pressure interaction. A sequence of such opto-mechanical interactions is utilized to exchange the information of the massive oscillator onto the optical field. Thus, any deformation of the commutation relation will induce a measurable deviation in the optical phase with respect to that of the usual canonical commutation relation. By utilizing the conventional interferometric techniques, it is possible to provide a high precision measurement of this optical phase shift and, thus, it is viable to shed light on the effects of noncommutative theories at the Planck-scale regime.   
\section{Opto-mechanical scheme for noncommutative systems}\label{sec2}
In a recent study \cite{Liang}, the author explored an experiment to test noncommutative theories based on the Aharonov-Bohm effect in nano-scale quantum mechanics. Here we use an opto-mechanical scheme to test such theories. The mechanical oscillator that we shall consider here are based on a two-dimensional version of the general form of noncommutative space introduced in \myref{NCOM}. For simplicity we take the dimensionless position and momentum operators of the oscillator
\begin{alignat}{1}\label{Quadrature}
& X_\text{m}=x\sqrt{\frac{m\omega_\text{m}}{\hbar}}, \quad Y_\text{m}=y\sqrt{\frac{m\omega_\text{m}}{\hbar}}, \\
& P_\text{m}^X=\frac{p_x}{\sqrt{\hbar m \omega_\text{m}}}, \quad P_\text{m}^Y=\frac{p_y}{\sqrt{\hbar m \omega_\text{m}}}, \notag
\end{alignat}
which are familiar as the quadrature operators in the literature. We also assume that the oscillator is isotropic, so that $\omega_\text{m}^X=\omega_\text{m}^Y=\omega_\text{m}$. The quadratures described in \myref{Quadrature}, therefore, satisfy the following commutation relations
\begin{alignat}{1}
[X_\text{m},Y_\text{m}]&=\frac{m\omega_\text{m}}{\hbar}[x,y],~[P_\text{m}^X,P_\text{m}^Y]=\frac{1}{\hbar m\omega_\text{m}}[p_x,p_y] \notag \\
& [X_\text{m},P_\text{m}^X]=[Y_\text{m},P_\text{m}^Y]=\frac{1}{\hbar}[x,p_x].
\end{alignat}
The opto-mechanical scheme is realized by means of an interaction between an optical pulse and a mechanical oscillator of mass $m$ and angular frequency $\omega_\text{m}$. In order to understand this interaction, we shall utilize a unitary displacement operator that displaces the quadratures of a massive mechanical oscillator in phase space induced by the optical field of interaction length $\boldsymbol{\lambda}=(\lambda_1,\lambda_2)$
\begin{equation}
U_{\text{LM}}=e^{in_L(\lambda_1 X_\text{m}+\lambda_2Y_\text{m})},
\end{equation}
with $n_\text{L}$ being the photon number operator. Here the subscripts M and L stands for the identification of the mechanical quantities and the optical degrees of freedom, respectively. A sequence of four such radiation pressure interactions effectively forms an optical cavity (resonator) yielding the total interaction operator
\begin{alignat}{1}\label{TotalInt}
\xi=& e^{in_L(\lambda_1 P_\text{m}^X+\lambda_2P_\text{m}^Y)} e^{-in_L(\lambda_1 X_\text{m}+\lambda_2Y_\text{m})} \\
&\times e^{-in_L(\lambda_1 P_\text{m}^X+\lambda_2P_\text{m}^Y)} e^{in_L(\lambda_1 X_\text{m}+\lambda_2Y_\text{m})}, \notag
\end{alignat}
which displaces the mechanical state around a complete loop in phase space. This displacement creates additional phase to the state, which can be utilized to engineer quantum gates \cite{Sorensen,Milburn,Leibfried}. The displacement operator \myref{TotalInt} contains all the informations of the mechanical oscillator in terms of the quadrature operators and, our work is to transmit this mechanical information onto the optical field. Thus, the phase shift created by the mechanical oscillator will change the optical field accordingly, which  is easily measurable in a table top experiment. By quantifying this change in phase of the optical field, it is feasible to understand the effects arising from the noncommutative mechanical oscillator. What remains is to understand the change in the optical field. In order to compute this change we analyze the mean of the optical field operator
\begin{equation}\label{FO}
\langle a\rangle=\langle\alpha |\xi^\dagger a\xi|\alpha\rangle=\langle a\rangle_{_\text{QM}}e^{-i\Theta},
\end{equation}
where $a$ is the annihilation operator of the optical field $|\alpha\rangle$ is the input optical coherent state with mean photon number $N_p$. Here, $\Theta$ describes the additional phase on the light arising due to the presence of the noncommutative oscillator. The mean of the field operator using the standard quantum mechanics corresponding to a two-dimensional mechanical oscillator can be computed as \cite{Bosso}
\begin{equation}\label{FOQM}
\langle a\rangle_{_\text{QM}}=\alpha e^{-i(\lambda_1^2+\lambda_2^2)-N_p(1-e^{-2i(\lambda_1^2+\lambda_2^2)})}.
\end{equation}
In order to compute the phase $\Theta$, let us first simplify the interaction operator \myref{TotalInt}. A straightforward calculation with the help of the Baker-Campbell-Hausdorff formula $\log(e^Ae^B)=A+B+[A,B]/2+([A,[A,B]]+[B,[B,A]])/12-[B,[A,[A,B]]]/24-....$, we obtain a simple form of \myref{TotalInt} as given by
\begin{equation}\label{TotalIntSim}
\xi=e^{-i\left(1+\frac{\tilde{\theta}\tilde{\Omega}}{4\hbar^2}\right)(\lambda_1^2+\lambda_2^2)n_L^2}.
\end{equation}
Note that, in the limit $\lambda_1=\lambda,\lambda_2=\theta=\Omega=0$, the opto-mechanical interaction operator \myref{TotalIntSim} reduces to that of the ordinary quantum mechanics \cite{Bosso}. After completing $N$ similar cycles inside the cavity, the total interaction operator for the optical pulse \myref{TotalIntSim} turns out to be
\begin{equation}
\xi^N=e^{-iN\left(1+\frac{\tilde{\theta}\tilde{\Omega}}{4\hbar^2}\right)(\lambda_1^2+\lambda_2^2)n_L^2},
\end{equation}
from which we calculate
\begin{equation}
\left(\xi^N\right)^\dagger a\xi^N=e^{-2iN\lambda^2(1+\frac{\tilde{\theta}\tilde{\Omega}}{4\hbar^2})(2n_L+1)}a,
\end{equation}
where we have considered $\lambda_1=\lambda_2=\lambda$, since the oscillator is isotropic. It should be noted that it is not necessary to consider the $N$ number of cycling of the optical pulse, however, as indicated in \cite{Bosso} that considering N number of cycles one can effectively reduce the background noise of the system, which will be helpful to increase the efficiency of the system. Nevertheless, by using the following well-known properties of coherent states
\begin{alignat}{1}
& a|\alpha\rangle=\alpha|\alpha\rangle, \quad e^{i\phi n_L}|\alpha\rangle=|\alpha e^{i\phi}\rangle, \\
& \langle\alpha|\beta\rangle =e^{\alpha^\ast\beta-\frac{|\alpha|^2+|\beta|^2}{2}},~\langle n_L\rangle =\langle\alpha|n_L|\alpha\rangle=|\alpha|^2,
\end{alignat}
we compute the average of the optical field operator after $N$ cycles
\begin{alignat}{1}
\langle a\rangle_{_N} &:= \left\langle\alpha\left\vert\left(\xi^N\right)^\dagger a\xi^N\right\vert\alpha\right\rangle = \langle \alpha\rangle_{_{\text{QM},N}}e^{-i\Theta (N)} \label{FON} \\
&=\alpha e^{-2iN\lambda^2(1+\frac{\tilde{\theta}\tilde{\Omega}}{4\hbar^2})-N_p\left(1-e^{-4iN\lambda^2(1+\frac{\tilde{\theta}\tilde{\Omega}}{4\hbar^2})}\right)}. \label{FON1}
\end{alignat}
By comparing \myref{FON} and \myref{FON1} along with the help of \myref{FOQM}, we finally obtain the desired additional phase
\begin{alignat}{1}
\Theta (N) &=\frac{N\lambda^2\tilde{\theta}\tilde{\Omega}}{2\hbar^2}+iN_pe^{-4iN\lambda^2}\left(e^{\frac{iN\lambda^2\tilde{\theta}\tilde{\Omega}}{\hbar^2}}-1\right), \label{Theta} \\
\big|\Theta (N)\big| &=\frac{N\lambda^2\tilde{\theta}\tilde{\Omega}}{2\hbar^2}+2N_p\sin\frac{N\lambda^2\tilde{\theta}\tilde{\Omega}}{2\hbar^2}. \label{ModTheta}
\end{alignat}
Notice that the parameters $\tilde{\theta}$ and $\tilde{\Omega}$ always appear as a product between them in the expression of measurable quantities like $\Theta(N)$ in \myref{Theta} and \myref{ModTheta} and, thus, the actual measurement may not be able to disentangle the two parameters. This may be considered as a drawback of our procedure, however, it does not affect the the original process of measurement anyway. It should also be mentioned that in \cite{Pikovski}, the authors computed an approximate expression of $\Theta$ coming from a different system, which is in the context of generalized uncertainty principle. Here we have an exact expression  \myref{Theta} instead.
\begin{figure}
\includegraphics[scale=0.33]{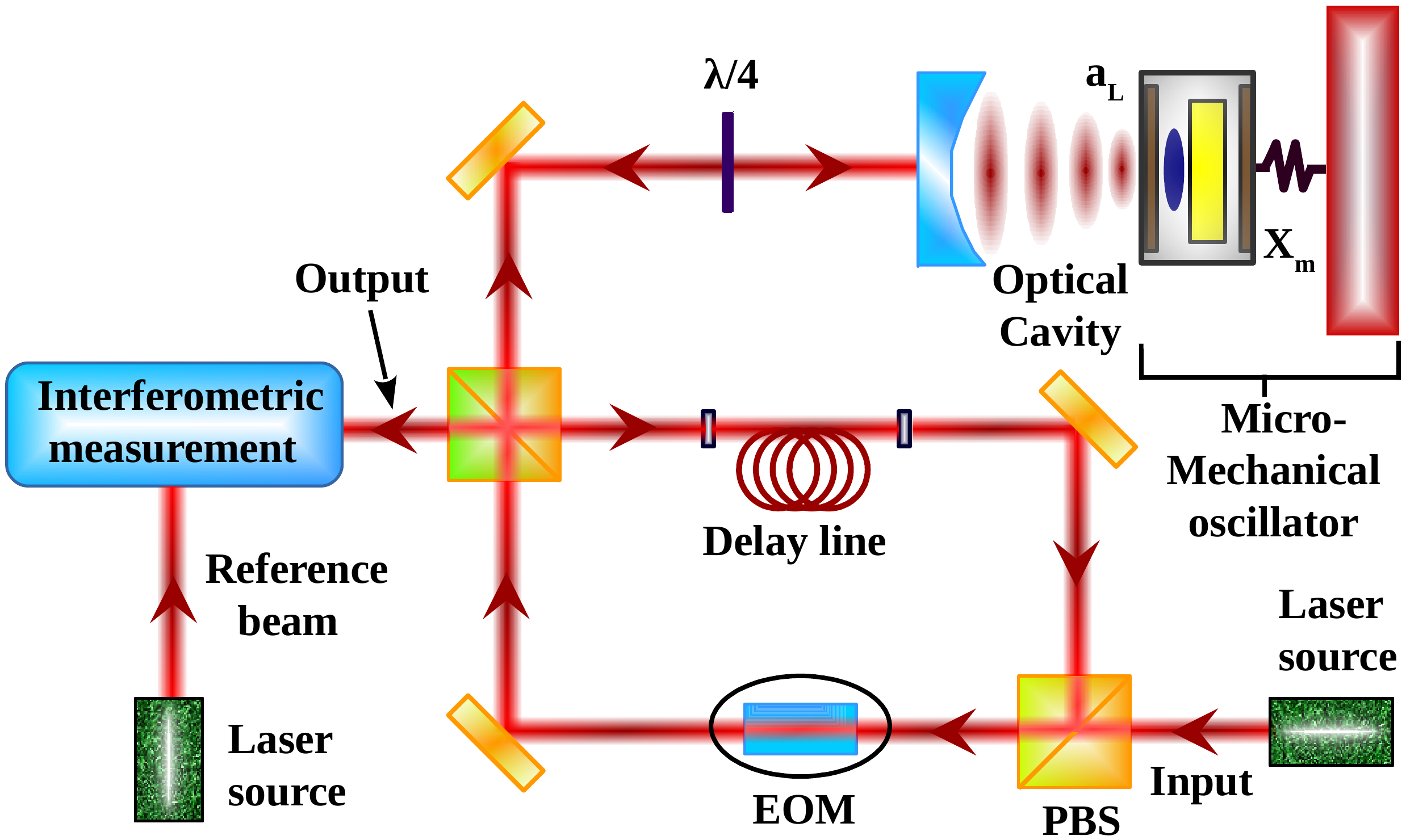}
\caption{Proposed experimental setup to probe the effects of noncommutative structure.}
\label{Fig1}
\end{figure}
\section{Experimental realization}\label{sec3}  
Let us now discuss a realistic experimental scenario that can measure the additional phase \myref{ModTheta} arising due to the noncommutative structure. This measurement will in turn prove the existence of the noncommutative structure of space-time and, thus, will probe the information about the high energy regime relevant to quantum gravity. The opto-mechanical experimental scheme that we shall consider here has been shown to be useful in different phenomena \cite{Kippenberg,Connell,Teufel, Grossardt,Gan} including a case where the existence of a minimal length scale has been explored in the framework of generalized uncertainty principle \cite{Pikovski}. The schematic of the experimental set-up as shown in Fig. \ref{Fig1} is more or less similar in different scenarios, however,
\begin{figure}[b]
\includegraphics[scale=0.39]{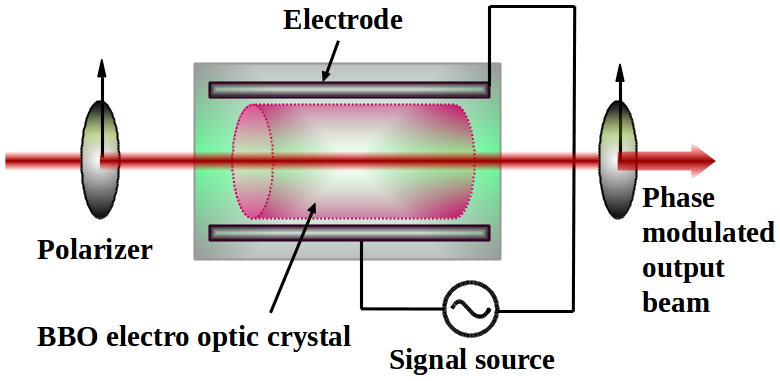}
\caption{Electro-optic modulator (EOM): It uses a particular type of electro-optic effect (Pockels effect) to modulate the phase of the incident beam of light. The EOM that we use here is a refractive modulator (whose refractive index changes due to the application of electric signal that causes the modulation) that produces an output beam of light whose phase is modulated with the electric signal applied to the Beta Barium borate (BBO) electro-optic crystal. The modulation can be controlled by the source of electric signal according to our requirement.}
\label{Fig2}
\end{figure}
the principle changes depending on the requirement. In our case, the device couples a micro-mechanical oscillator with an optical pulse via radiation pressure inside a high-finesse optical cavity. More specifically, a polarized phase-modulated optical beam is prepared by transmitting an incident optical pulse, labeled by ``input", through a polarizing beam splitter (PBS) followed by an electro-optic modulator (EOM) (Fig. \ref{Fig2}). Then, it is allowed to interact with a mechanical oscillator having position $X_m$ via a cavity field $a_L$. Afterwards, it reflects back towards the source with a negligible amount of scattering and enters into a delay line during which time the mechanical resonator evolves one quarter of a mechanical period. The optical pulse is now vertically polarized and becomes horizontal polarized after passing it through the EOM and, thus, is allowed to interact again with the mechanical resonator. The same process is repeated four times in total so that the canonical commutator is mapped onto the optical field as described in \myref{TotalInt}. Finally the EOM is operated in such a way that it does not rotate the polarization and, therefore, the optical pulse exits the system through the way labeled by ``output". If the entire process is repeated for more times (say $N$ times), the experimental error proportionally goes down as indicated in \cite{Bosso}. Nevertheless, the output coherent beam can now be studied by using an interferometric system, as described in Fig. \ref{Fig3}. A precise measurement of interference fringe shifts caused by the output beam with respect to a reference beam associated with the usual oscillator can be carried out. Thus, the deviation of the phase of the light field \myref{ModTheta} which occurs due to the presence of noncommutative structure can be measured with the given experimental setup with a very high accuracy.
\begin{figure}
\includegraphics[scale=0.52]{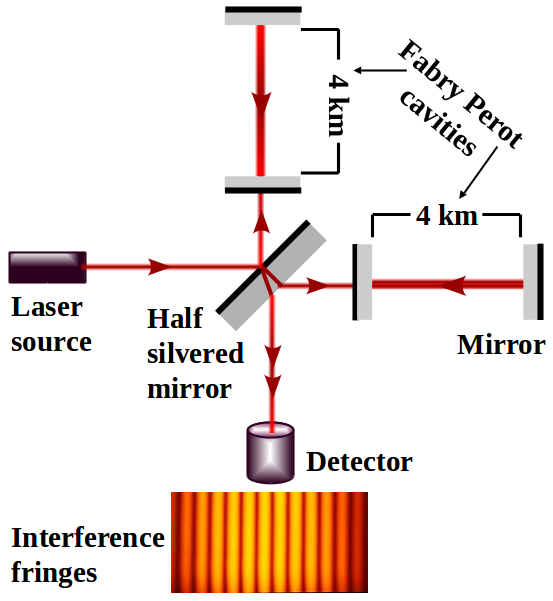}
\caption{Schematic illustration of a Michelson interferometer. It consists of a half-silvered mirror, two Fabry-Perot cavities, a laser source and a detector. The Fabry-Perot cavity is designed in a way where two flat mirrors are positioned normally to the direction of propagation of the two waves. The incident light into the cavity is accomplished by multiple reflections between the two-mirrors which creates the interference condition.}
\label{Fig3}
\end{figure}

The schematic of the experimental setup as shown in Fig. \ref{Fig1} is motivated mainly from the earlier work \cite{Pikovski}, however, it should be noted that our experimental scheme is different than that reported in \cite{Pikovski}, in many ways. In fact, we introduce some new features which eventually improves the efficiency of the experiment. Firstly, the EOM is composed of a BBO electro-optic crystal and is controlled by the electric signal and, thus, the EOM behaves as a refractive phase modulator. It is designed in such a way that the electrodes are free from any external and internal spring transduction and, therefore, they are automatically adjusted due to a slight phase difference created inside the EOM. The device being made of piezoelectric material and being controlled by the electric signal, the accuracy of the system is improved to quite a good extent. Secondly, unlike the usual mechanical oscillator as used in \cite{Pikovski}, we propose a micro-mechanical oscillator consisting of plate spring system as shown in Fig. \ref{Fig1}. One of the advantages of using such oscillators is that it vibrates at a high frequency \cite{Candler}, and is exploited in many highly efficient experiments including signal filtration, mass and motion sensing \cite{Nguyen}, etc.

Thus, our experimental setup is no longer attributed to an ordinary opto-mechanical setup, rather it can be referred as a Micro-Electro-Mechanical System (MEMS) based oscillator, which is a more efficient version of the usual opto-mechanical systems. The oscillator comprises of frequency selective material called the micro-mechanical resonator and a feedback amplifier which sustains the resonance in the system by transferring the electrical energy into mechanical energy \cite{Lam}. The basic principle of transduction in the MEMS oscillator proposed here is piezoelectric \cite{Franke}. The advantages of using such oscillator unit are that it keeps track of real time as well as sets the clocking of logic circuits and maintains a periodic frequency for data transmission. MEMS based micro-mechanical oscillators have superior frequency stability. Another major advantage of using a MEMS oscillator assembly is the presence of high resonator Q-factor (which is proportional to low resonator impedance), which ensures the narrow band filtering and less amount of noise. The MEMS scale integration of the oscillator assembly is an advantage to the detection system because the components like resonator, its intermittent electric circuitry and the position of plate mass spring system is done on a single wafer. The series of fabrication steps eliminates the time consumption and sequential implantation of components which at times might deflect and cause fault lines. The system used in \cite{Pikovski} does not consider the settings that are discussed above and, therefore, their system may be volatile to the losses and, hence, the process of quantum scale probing may be affected drastically.  

We will now discuss the feasibility of the proposed experiment, i.e. whether the scheme can be implemented in the laboratory with the resources available to us. The opto-mechanical interaction can be described by the inter-cavity Hamiltonian \cite{Pikovski} $H=\hbar(\omega_mn_m-g_0n_LX_m-h_0n_LY_m)$, with $n_m$ being the mechanical number operator and $g_0=h_0=\omega_c\sqrt{\hbar}/(L\sqrt{m\omega_m})$ being the opto-mechanical coupling rate with the mean cavity frequency $\omega_c$ and mean cavity length $L$. It should be mentioned that $g_0$ and $h_0$ would not be equal to each other if we would not consider the oscillator to be isotropic. However, in our case, it is isotropic and, therefore, in this scenario, the effective interaction length becomes \cite{Pikovski} $\lambda=(\lambda_1+\lambda_2)/2\simeq (g_0+h_0)/2\kappa=4\mathcal{F}\sqrt{\hbar}/(\lambda_L\sqrt{m\omega_m})$, where $\mathcal{F}$ is the cavity finesse, $\kappa$ is the optical amplitude decay rate and $\lambda_L$ is the optical wavelength. Under this framework, the opto-mechanical phase \myref{ModTheta} is modified as
\begin{equation}
\big|\Theta (N)\big| =\gamma+2N_p\sin\gamma, \quad \gamma=\frac{8\theta\Omega N\hbar\mathcal{F}^2}{m\omega_m\lambda_L^2}.
\end{equation}
In a real life experiment there are many sources for background noise that may limit our ability to detect the phase in the laboratory. Therefore, it becomes necessary to study the signal to noise ratio (SNR) $=\Theta/\delta\varphi$, with $\delta\varphi$ being the uncertainty in measuring the optical phase shift $\varphi$. In a realistic scenario it is desirable that SNR $>1$. An ideal experiment with coherent state of light with mean photon number $N_p$ yields the phase uncertainty $\delta\varphi=1/\sqrt{N_pN_r}$, where $N_r$ is the number of independent runs of the experiment. In addition, the uncertainty to the scaling of the signal should increase with number of loops $N$. Therefore, in ideal situation SNR $=N\sqrt{N_pN_r}$. If we fix the experimental parameters as $\mathcal{F}=0.1,~m=10^{-7}~\text{kg},~N_p=10^6,~\omega_m/2\pi=10^5~\text{Hz},~\lambda_L=1064~\text{nm},~N_r=10^2,~N=1$ , all of which are in the range of current experiments \cite{Corbitt,Thompson, Groblacher,Verlot,Kleckner}, we obtain $|\Theta(1)|\sim 10^{-4},~\delta\varphi=10^{-4}$ and, thus, the SNR $\sim 1$. Here the parameters are chosen such that a precision of $\theta\Omega\sim 10^{12}$ can be achieved, which amounts to measuring the deformations due to the noncommutative system upto which the noncommutative parameters are found to be bounded currently ($\tilde{\theta}\sim 10^{-8}~\text{GeV}^{-2}$). However, a more interesting effect follows from the case when we increase the cavity finesse $\mathcal{F}$ to $10^5$ by keeping all other parameters same as before. In this case, we are able to reach the accuracy $\delta\theta\sim 1$ and $\delta\Omega\sim 1$ (with SNR $\sim 1$) for which the minimal length turns out to be of the order of the Planck length. It should be noted that the modified value of the cavity finesse is also within the reach of current technology \cite{Corbitt,Thompson, Groblacher,Verlot,Kleckner}. 

Therefore, it turns out that it is possible to reach even to the Planck scale for which one only needs to improve the finesse of the cavity, which is proportional to the quality factor. This is surely consistent and desirable that in order to be able to probe a higher accuracy one must require the improvement of the quality factor of the cavity, which is defined as the ability of the cavity to confine the field. Of course, there is the flexibility in choosing the experimental parameters $\mathcal{F},m,N_p,\omega_m,\lambda_L,N_r,N$ in some other ways in order to achieve the experimental goals, but we provide a possible example here. In summary, our setup may be able to detect any level of accuracy in the length measurement in between the current lower bound and the Planck length. So, if the theoretical study improves the current lower bound in near future, the present experimental setup can be utilized for the detection of the corresponding length scale with a suitable adjustment of the experimental parameters. Thus, it is obvious that the proposed scheme could offer a feasible route to probe the possible effects of the noncommutative structure and, hence, of quantum gravity in a table-top experiment.
\section{Conclusions}\label{sec4} 
We have explored a detailed procedure to realize the effects of noncommutative theories in laboratory, which in turn could lead us towards the proper route to quantum theory of gravity. Our method utilizes an opto-mechanical setup that helps us to transfer the information of a noncommutative oscillator to the high intensity optical pulse in terms of a sequence of opto-mechanical interaction inside an optical resonator. Consequently, we end up with an optical phase shift that is easily measurable with a very high accuracy through an interferometric system. This makes the whole procedure much easier to collect the informations of noncommutative structures thorough an elegant optical system already available to us. Thus, it may bypass all the difficulties of probing high energy scales through scattering experiments.     

\vspace{0.5cm} \noindent \textbf{Acknowledgements:} The authors would like to thank Saurya Das for useful discussions. SD is supported by an INSPIRE Faculty Grant by the Department of Science and Technology, Government of India. AB is supported by MHRD, Government of India and would like to thank Department of Physics and MMED at NIT Srinagar for carrying out her research pursuit.



\end{document}